\documentclass[twocolumn,prl,preprintnumbers]{revtex4}

\usepackage[dvips]{color}

\usepackage{times}
\usepackage{graphicx}
\usepackage{dcolumn}
\usepackage{bm}

\begin{document}


\title{\sf \Huge Encoding folding paths of RNA switches}

\author{A.~Xayaphoummine$^{\rm ~1}$, V.~Viasnoff$^{\rm ~2}$, S.~Harlepp$^{\rm
 ~1}$ and H.~Isambert$^{\rm ~1,2,*}$}

\vskip 0.2cm

\affiliation{\vskip 0.2cm
${\rm ~1}$ Laboratoire de Dynamique des Fluides Complexes, CNRS-ULP, Institut
de Physique, 3 rue de l'Universit\'e, 67000 Strasbourg, France \rm}%

\affiliation{\vskip 0.2cm
${\rm ~2}$ \it RNA dynamics and biomolecular systems, Physico-chimie Curie, CNRS UMR168, Institut Curie, Section de
Recherche, 11 rue P. \& M. Curie, 75005 Paris, France}%

\maketitle



\small

\noindent
{\bf 
RNA co-transcriptional folding has long been suspected to play an active role in helping proper native folding of ribozymes and structured regulatory motifs in mRNA untranslated regions. Yet, the underlying mechanisms and coding requirements for efficient co-transcriptional folding remain unclear. Traditional approaches have intrinsic limitations to dissect RNA folding paths, as they rely on sequence mutations or circular permutations that typically perturb both RNA folding paths {\em and}  equilibrium structures. Here, we show that exploiting sequence symmetries instead of mutations can circumvent this problem by essentially decoupling folding paths from equilibrium structures of designed RNA sequences. Using bistable RNA switches with symmetrical helices conserved under sequence reversal, we demonstrate experimentally that native and transiently formed helices can guide efficient co-transcriptional folding into either long-lived structure of these RNA switches. Their folding path is controlled by the {\em order} of helix nucleations and subsequent exchanges during transcription, and may also be redirected by transient antisense interactions. Hence, transient intra- and intermolecular base pair interactions can effectively regulate the folding of nascent RNA molecules into different native structures, provided limited coding requirements, as discussed from an information theory perspective. This constitutive coupling between RNA synthesis and RNA folding regulation may have enabled the early emergence of autonomous RNA-based regulation networks.\\ 
}

\noindent
{\large \bf Introduction}

RNA molecules exhibit a wide range of functions from essential components 
of the transcription/translation machinery\cite{dahlberg} to natural or 
{\em in vitro} selected 
ribozymes\cite{cech,bartelszostak} or  aptamers\cite{joyce,ellington,tuerk}
and different classes of gene expression regulators ({{\em e.g.}} miRNA,
siRNA, riboswitches)\cite{mironov,nahvi,winkler,bartel,nudler,hannon,breaker}. 
\textcolor [rgb]{0,0,0}{The functional control of many of these RNA molecules 
hinges on the formation of specific base pairs in  cis or trans and secondary
structure rearrangements between long-lived alternative folds.  
Yet, because of their limited 4-letter alphabet and strong base pair
stacking energies, RNAs are also prone to adopt long-lived  misfolded
structures\cite{uhlenbeck}, as observed for instance upon heat renaturation.
Hence, efficient RNA folding paths leading to properly folded structures
bear an important role in the regulatory function of non-coding RNAs and mRNA
 untranslated regions\cite{uhlenbeck}.} 

\textcolor [rgb]{0,0,0}{It has long been proposed\cite{boyle,kramer,nussinov}
  that,} during
transcription, the progressive folding of nascent RNAs limits the number of folding pathways,
 presumably facilitating their rapid folding into proper native structures.
\textcolor [rgb]{0,0,0}{It is not clear, however,
whether native domains fold sequentially and independently from one another or
whether co-transcriptional folding 
paths result from more intricate interactions between domains and individual
helices.}  
\textcolor [rgb]{0,0,0}{Transcriptional RNA switches provide interesting
  examples of co-transcriptional folding paths with local competition
  between newly formed and alternative helices.} 
Such natural RNA switches are typically found in virus or plasmid
genomes\cite{biebricher,crothers,repsilber,gultyaev1998,olsthoorn} and in
bacterial mRNA untranslated regions where they regulate
gene expression at the level of transcription elongation ({\it e.g.} through
termination/antitermination mechanism)\cite{putzer1,putzer,wickiser} or at the level of translation
initiation ({\it e.g.} through sequestration of Shine-Dalgarno motifs)\cite{brunel2,romby,caillet,vanmeerten,smit,moller-jensen}. 
The structural changes controlling their regulatory function may correspond
to a switch in equilibrium structure or in 
co-transcriptional folding path caused by binding an effector ({\it e.g.} a
protein, a small metabolite or an antisense sequence)\cite{soukup}.  
Alternatively, RNA switches may operate through spontaneous or assisted
relaxation 
of an initially metastable co-transcriptional fold\cite{nagel2002}. Hence, RNA
switches can have stringent needs to control their folding between alternative
structural folds, which makes them ideal candidates to dissect RNA
co-transcriptional folding mechanisms and estimate the minimal  
\textcolor [rgb]{0,0,0}{sequence constraints to encode them}.

Several inspiring reports have demonstrated the importance of
co-transcriptional
folding\cite{repsilber,groeneveld,gultyaev1995,poot,gerdes1997,gultyaev1997,franch1997,nagel1999}, 
and folding pathways of {\em E. coli}
RNaseP RNA\cite{pan1999,pan2} and {\em Tetrahymena} group I
intron\cite{woodson,heilman1} have been probed using circularly permutated
variants of their wild type sequences (see Discussion).
Yet, dissecting folding paths of natural RNAs remains generally difficult due
to  two fundamental issues:
{\it i)} sequence mutations or circular permutations  generally 
affect both RNA  folding paths {\em and} equilibrium structures (hence 
preventing independent probing of folding paths on their own), and 
{\it ii)} many natural non-coding RNAs have likely evolved to perform 
multiple interdependent functions, which are all encoded 
on their primary sequence and thereby all
potentially affected, directly or indirectly, by sequence mutations.

To circumvent these limitations, we propose to use artificial RNA
switches, presumably void of biological functions, and investigate how to
efficiently encode their folding paths by exploiting simple
sequence symmetries, instead of extensive (and possibly non-conclusive)
mutation studies. 
Beyond specific examples of natural or designed RNA sequences, we aim at
delineating {\em general} mechanisms and coding requirements  
for efficient co-transcriptional folding paths. 

In a nutshell, we have designed a pair of synthetic RNA switches 
\textcolor [rgb]{0,0,0}{sharing strong sequence symmetries, so that both
  molecules partition, at equilibrium, into equivalent branched and rod-like
  nested structures with nearly the same free energy.} 
Yet, in spite of this structural equivalence between the two RNA switches at
equilibrium, we demonstrate that their  folding path can be encoded 
to {\it guide} the first RNA switch exclusively into the branched structure
while the other switch adopts instead the rod-like 
nested structure by the end of transcription. 
\textcolor [rgb]{0,0,0}{This shows that
folding paths do not simply result from the sequential formation of
native helices in their order of appearance during transcription ({\it i.e.}
`sequential folding', see Discussion). 
Instead, efficient folding paths rely on  the  relative
stability between native and non-native 
helices  together with their 
precise positional order
along the 5'-3' oriented sequence ({\it i.e.} `encoded
co-transcriptional folding').}
Furthermore, we show that efficient folding path can be redirected through 
transient antisense
interaction during transcription, suggesting 
an intrinsic and possibly ancestral coupling between RNA
synthesis and folding regulation.

\vspace{0.5cm}
\noindent
{\large \bf Materials and Methods}

\vspace{0.2cm} \noindent {\bf \em RNA switch design.}\\
RNA switches with encoded folding paths depicted in Figs.~1\&2 were designed 
using Kinefold online
server\cite{pnas2003,kinefold} ({\sf http://kinefold.curie.fr}). A starting GG doublet was chosen to ensure
efficient T7 transcription. 
The sequence of the ``direct'' RNA switch 
({\it i.e.} 5'-ABCD-3', 73~nucleotides, Fig.2) is: 
5'-GGAA-CCGUCUCCCUCUGCCAAAAGGUAGAGGGAGAUGGAGC-AUCUCUCUCUACGAAGCAGAGAGAGACGAAGG-3'. 
The ``reverse'' RNA switch has exactly the opposite sequence (or
reversed orientation), {\it i.e.} 5'-DCBA-3'. 
A ``reverse'' sequence with a {\it single} mutation U$_{38}$/C$_{38}$ 
was also studied to unambiguously establish 
the correspondence between branched versus rod-like structures and 
the two migrating bands on \textcolor[rgb]{0,0,0}{nondenaturing}  polyacrylamide gels, Fig.~3; It is:
5'-GGAAGCAGAGAGAGACGA-AGCAUCUCUCUCUACGAGGC$\!_{38}$AGAGGGAGAUGGAAAA-CCGUCUCCCUCUGCCAAGG-3'.
Complementary DNA oligonucleotides including T7 promoter and KpnI/StuI/BamHI
restriction sites at sequence extremities were bought from IBA-Naps, Germany.

\vspace{0.2cm} \noindent {\bf \em Sequence cloning and in vitro transcription.}\\
Sequences were inserted into pUC19 plasmid (between KpnI and BamH1 restriction
sites) using enzyme removal kits (Qiagen) and cloned into calcium
competent {\em E. coli} (DH5$\alpha$ strain) following standard cloning
protocoles. 
Following plasmid extraction (Genomed kit), inserts
were sequenced and cut at the StuI blunt restriction site located at 
the end of the desired DNA template.
Run off transcription was performed {\em in vitro} 
using  T7 RNA polymerase (New England Biolabs) for upto 4-5
hours at 37$^\circ$C or 25$^\circ$C. 
\textcolor[rgb]{0,0,0}{Heat renaturation was performed  from 85$^\circ$C to
  room temperature  in about 10~min (starting from 95$^\circ$C gave the same
  results).} 
Renatured and co-transcriptional native structures were then separated on 12\%
19:1 acryl-bisacrylamide \textcolor[rgb]{0,0,0}{nondenaturing}  gels (1X TAE,
temperature $<$10$^\circ$C)  
and observed using ethidium bromide staining (0.1$\mu$g/$\mu$l); 
Ethidium bromide slightly re-equilibrates the molecule
equilibrium partition between branched and rodlike structures 
during heat renaturation
but has no measurable effect at room temperature on the
strongly biased partition between co-transcriptional native structures.   
Controls using denaturing gels (6\%, 19:1 acryl-bisacrylamide, RNA in
formamide and 8M urea, 50$^\circ$C) showed that $>$90\% of transcripts had the
expected run off transcription length. 
Virtually no other bands ({\it i.e.} $<$5\% of total transcript) were observed
on \textcolor[rgb]{0,0,0}{nondenaturing}  gels either ({\it i.e.} apart from
the single or double bands shown in Figs.~2-4). 

\begin{figure}
\includegraphics{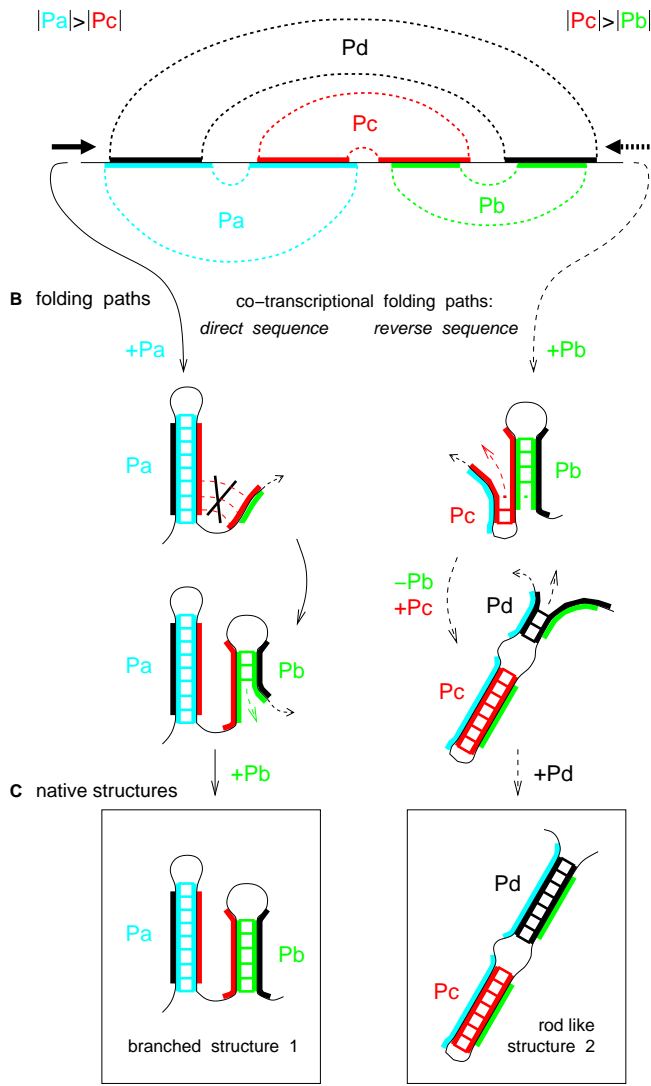}
\caption{\label{fig:wt} 
{\footnotesize
{\bf Encoded co-transcriptional folding path of a bistable RNA switch.}
{\bf A.} Bistable generic sequence with hierarchically overlapping helices.
{\bf B.} Opposite co-transcriptional folding paths for the direct and reverse
sequences rely on 
small asymmetries in helix
length in the direction of transcription ({\it i.e.} $\vert$Pa$\vert\!>\!\vert$Pc$\vert$ direct sequence and $\vert$Pc$\vert\!>\!\vert$Pb$\vert$ reverse sequence).
{\bf C.} {\it Either} branched {\it or} rod-like native structures are 
obtained depending
on the direction of transcription, although both structures can be designed to
co-exist at equilibrium. 
}}
\vspace{-.5cm}
\end{figure}

\begin{figure} 
\includegraphics{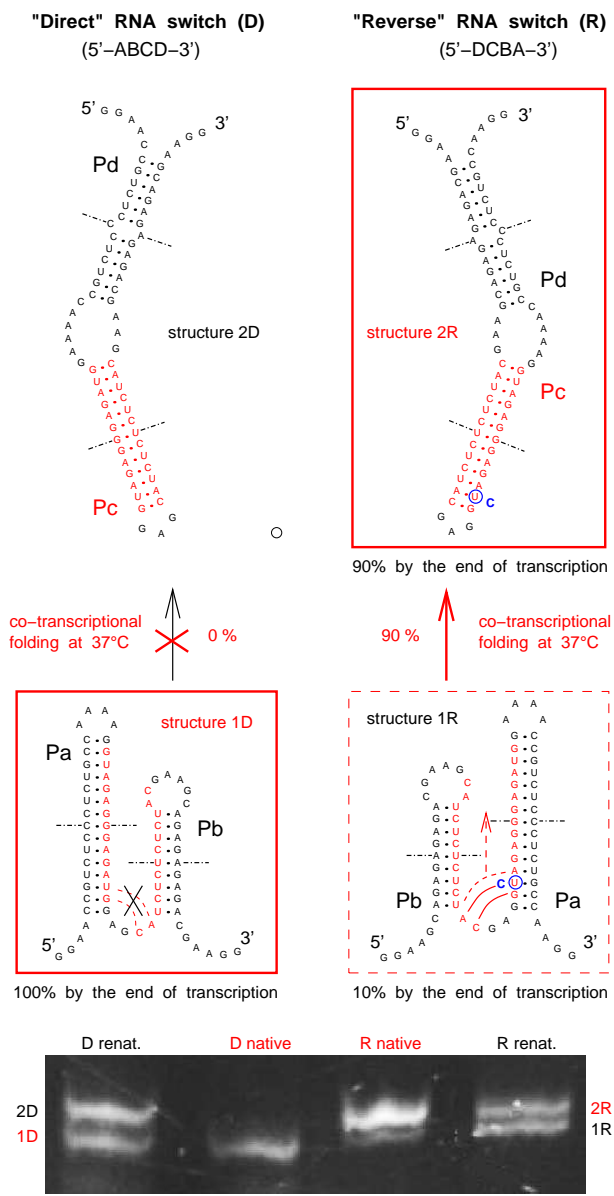}
\caption{\label{fig:wt} 
\footnotesize
{\bf Opposite co-transcriptional folding paths of a pair of RNA switches with
  `direct' and `reverse' sequences} ({\em i.e.} 5'-ABCD-3' vs
5'-DCBA-3'). Structures 1D and 1R (resp. 2D and 2R) of the direct and reverse
switches are energetically equivalent because of helix symmetries; dashed
lines indicate mirror symmetry of Pa, Pb, Pc and Pd which are therefore
conserved under sequence reversal relating direct and reverse
switches. Despite these strong similarities between D and R structures at
equilibrium, direct and reverse switches display `opposite' co-transcriptional
folding paths (direct switch into struct.1D and reverse switch into struct.2R)
guided through a helix encoded persistence (left) or exchange (right) during
{\it in vitro} transcription using T7 RNA polymerase (see Materials and Methods). 
} 
\vspace{-0.6cm}
\end{figure}

\vspace{0.5cm}
\noindent
{\large \bf Results}

\vspace{0.2cm}

\textcolor [rgb]{0,0,0}{The results section is organized into 
two complementary subsections.
The first one is primarily experimental and demonstrates, using sequence
symmetries, the basis for encoding efficient folding paths 
with a pair of `symmetrically equivalent' RNA switches 
  adopting {\em either} their branched {\em or}  rod-like structure by 
  the end of transcription.
The second subsection is  theoretical and discuss, from an information
  content perspective\cite{adami,szostak} and beyond sequence specific
  examples, the coding requirement for such efficient
  co-transcriptional folding.}

\vspace{0.2cm} 
\noindent {\bf \em Co-transcriptional folding of ``direct'' and ``reverse'' switches}

We decided to investigate the basic mechanisms and coding requirements for
efficient RNA folding paths with a stringent test case.
Following the  RNA switch design depicted on Fig.~1, 
we set out to encode {\em two} (oppositely oriented) folding paths on the 
{\em same} RNA sequence. 
The proposed bistable RNA switch 
should form  a branched structure (1) and a
 rod-like structure (2) with approximately the same free energy at
 equilibrium, and yet be guided into either one of these structures only, 
depending on the direction of synthesis\cite{nanomemories}, Fig.~1.

\begin{figure}
\includegraphics{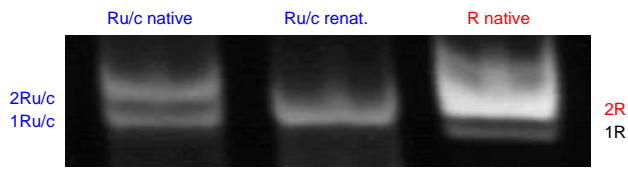}
\caption{\label{fig:wt} 
{\footnotesize {\bf Correspondence between branched versus rod-like structure
    and migrating bands.} 
A {\em single} mutation  
U$_{38}$/C$_{38}$ on the reverse sequence, Ru/c (see blue u/c mutation in
    Fig.~2)   
unambiguously demonstrates the correspondence
 between the stabilized branched structure and the lower band on the gel (see text).
}}
\vspace{-.5cm}
\end{figure}

In practice, however, 5'-3' vs 3'-5' folding paths cannot be probed
on the {\em same} RNA sequence, as there is no RNA polymerase known to perform
transcription in `opposite'  (3'-to-5') direction.
Hence, instead of studying a single RNA sequence, we have actually used a {\em
  pair} of RNA switches with {\it exactly}
opposite sequences , {\it i.e.} 5'-ABCD-3'
and 5'-DCBA-3' (see Materials and Methods).
It is important to note that, in general, such pairs of
RNA molecules do {\em not} adopt related structures at equilibrium, due to the
 large asymmetry between free energies of stacking base pairs with reversed
 orientation  ({\it e.g.} 5'-GC/GC-3'$\simeq$ -3.4~kcal/mol and
 3'-GC/GC-5'$\equiv$5'-CG/CG-3'$\simeq$ -2.4~kcal/mol). 
For this reason,
the pair of direct (D)
  and reverse (R) RNA switches, we have designed (Fig.~2), forms at equilibrium a branched structure (1) and a rod-like structure (2) constructed around
  {\em symmetric} helices, that are {\it exactly} conserved under sequence reversal (dashed lines on Fig.~2 indicate mirror symmetry of Pa,
   Pb, Pc and Pd helices). Thus, comparing transcription products of the direct and reverse sequences probes the directionality of their
  folding paths while keeping the equilibrium structures of both switches essentially equivalent by symmetry;
   since structures 1D and 1R (resp. 2D and 2R) of the direct and reverse switches are built on the {\em same} helices Pa and Pb (resp. Pc and Pd),
   their sole free energy difference concerns the small sequence dependent contribution of single stranded regions in the branched (resp. rod-like)
   structure (GNRA tetra-loops and all other sequence-dependent tabulated loops have been avoided).

\begin{figure}
\includegraphics{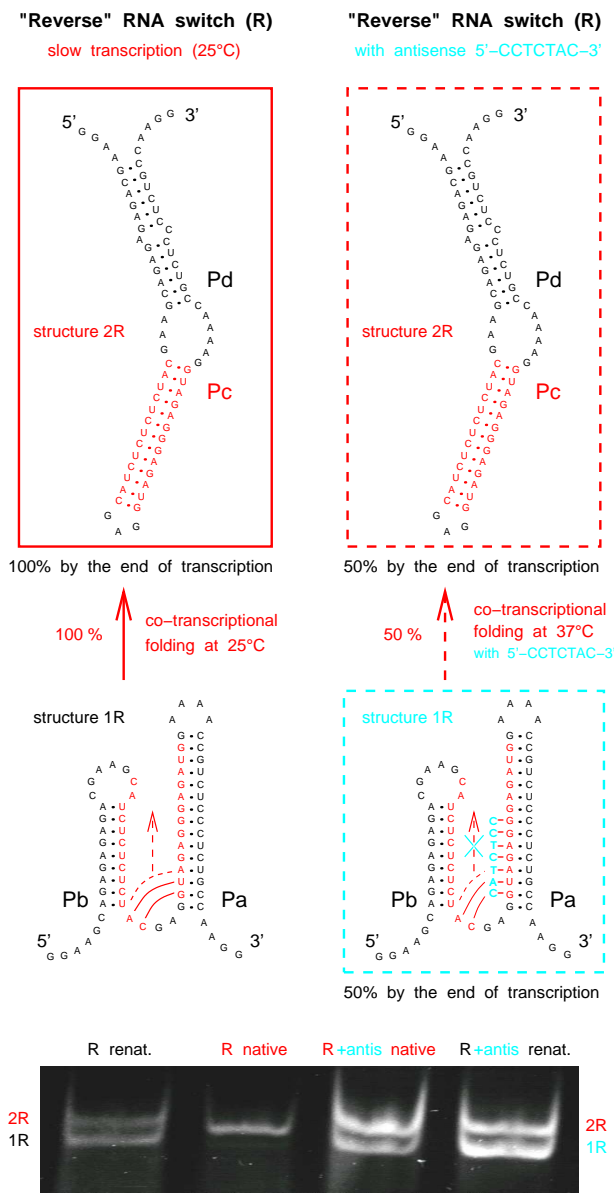}
\caption{\label{fig:wt} 
\footnotesize
{\bf Influence of temperature and transient antisense interactions on
  co-transcriptional folding.}
Equilibrium and native structures of reverse switch (R) with {\it in vitro} T7
transcription at 25$^\circ$C (left, see text) 
and
under {\it in vitro} T7 transcription in presence of 0.3~nmol/$\mu$l of the 7-nt
antisense DNA oligonucleotide CCTCTAC (right, see text).
Structures are separated on a 12\% 19:1 acryl-bisacrylamide
\textcolor[rgb]{0,0,0}{nondenaturing}  gel 
(temperature $<$10$^\circ$C) and observed using ethidium bromide staining as
on Fig.~2  (see Materials and Methods). 
}
\vspace{-.5cm}
\end{figure}
 
Fig.~2 demonstrates that, 
in spite of these strong similarities between equilibrium structures, 
the two RNA switches are indeed guided towards
 two distinct native structures upon {\it in vitro} transcription (see
 Materials and Methods).
The correspondence between branched versus rod-like structures and 
migrating bands on polyacrylamide gels was unambiguously  established using
 a {\em single} mutation  
U$_{38}$/C$_{38}$ on the reverse sequence, Fig.~2.  
This mutation stabilizes the branched structure (UG$>$CG) relative to the
 rod-like structure (AU$<$AC) at equilibrium, hence demonstrating the
 correspondence between branched structure and lower band, Fig.~3. 
Note, however, that this mutation also perturbs the co-transcriptional 
folding path by redirecting about half of the molecules
into the branched structure, hence illustrating the difficulty to dissect
 independently folding paths from equilibrium structures with sequence 
mutations only (see Introduction).

These results strongly
  support the co-transcriptional folding principles depicted on Fig.~1 which primarily rely on the difference in helix
  length ({\it i.e.} $\vert$Pa$\vert\!>\!\vert$Pc$\vert$ for the direct switch and $\vert$Pc$\vert\!>\!\vert$Pb$\vert$ for the reverse switch)
  to code for the co-transcriptional formation of structures 1D and 2R, respectively. This small asymmetry between successive overlapping helices
  in the direction of transcription induces a divergence in structural cascades between the two co-transcriptional folding paths; namely,
   the red helix Pc cannot displace and replace the longer (and stronger) helix Pa previously formed by the direct switch during transcription
   [$\vert$Pa-Pc$\vert/\vert$Pc$\vert\simeq$15\%], while Pc does displace and replace the shorter (and weaker) helix Pb initially formed by the
   nascent reverse sequence [$\vert$Pc-Pb$\vert/\vert$Pc$\vert\simeq$30\%]. The efficacy of these folding paths using T7 polymerase
   appears maximum ($\simeq$100\%) for the direct sequence while it is about 90\% for the reverse switch at 37$^\circ$C suggesting that the branch migration
   exchange between Pb and Pc is not always successful in these conditions (T7
  transcription rate is about 200-400~nt/s at 37$^\circ$C); however we found that
   the folding bifurcation is almost always achieved
 ($\simeq$100\%, Fig.~4) at lower temperature 25$^\circ$C which decreases 3 to
 4 folds T7 transcription rate\cite{chamberlin}.
\textcolor[rgb]{0,0,0}{This small improvement in bifurcation efficiencies
  suggests that the decreasing elongation rate indeed prevails over other
  opposite kinetic factors at lower temperature. In particular, nucleation of
  Pc (which probably 
  involves the opening of one or two   base pairs of Pb, as sketched in Fig.~2)
  and the subsequent branch migration between Pb and Pc are probably both
  slowed down at lower temperature.}  

Overall, this demonstrates that the competition between native and non-native
helices can lead to efficient co-transcriptional folding paths of RNA switches 
{\it independently} from their actual equilibrium structures.

Moreover, we found that the folding path of the reverse switch could be
significantly redirected towards structure 1R ($\simeq$50\%) 
through {\it transient} antisense interactions, Fig.~4 (right). This is simply
achieved using a 7-nt-long antisense oligonucleotide designed to interfere, 
through competing interactions, with the encoded exchange between helices Pb
and Pc, Fig.~5. Note, in particular, that the hybridized
 antisense probe is eventually displaced by the longer (and stronger)
 downstream helix Pa of each nascent reverse sequence (no global
 shift of equilibrium bands is observed between transcriptional folds 
formed in presence or absence of antisense probe, Fig.~4). 
This shows that  transient antisense interactions can, in principle, control
multiple turnovers of redirected folding pathways.

 Hence, transient intra- and intermolecular base pair interactions can efficiently {\it regulate} the folding of nascent RNA molecules between
 alternative long-lived native structures, irrespective of their actual thermodynamic stability. Indeed, once formed, the co-transcriptional structures
  1D and 2R remain trapped out-of-equilibrium for more than a day at room temperature (data not shown) demonstrating that these RNA switches can reliably
  store information on physiological time scales with their
 co-transcriptionally folded structures. 
In another context, the ability to control folding between distinct long-lived 
structures of nucleic acids using electrical\cite{nanomemories} or 
thermal\cite{nanoletter} stimuli, instead of transcription, could also lead 
to nanotechnology applications.

While our conclusions are based on particular examples of synthetic RNA switches related by sequence reversal and helix symmetries, we want
to stress that these strong symmetry constraints are solely instrumental in demonstrating the possible independence between encoded
 folding paths and low free energy RNA structures. These symmetries are not directly used nor necessary to achieve efficient co-transcriptional folding. On the contrary, imposing such strong sequence symmetries greatly
 limits the  additional ``information content'' that can possibly be encoded on
 the sequence. In the next subsection, we discuss how this use of sequence
 symmetries can actually be formalized to provide quantitative estimates on
 the minimum coding requirement for selective folding paths
 of  {\it generic} RNA swiches. 

\begin{figure}
\includegraphics{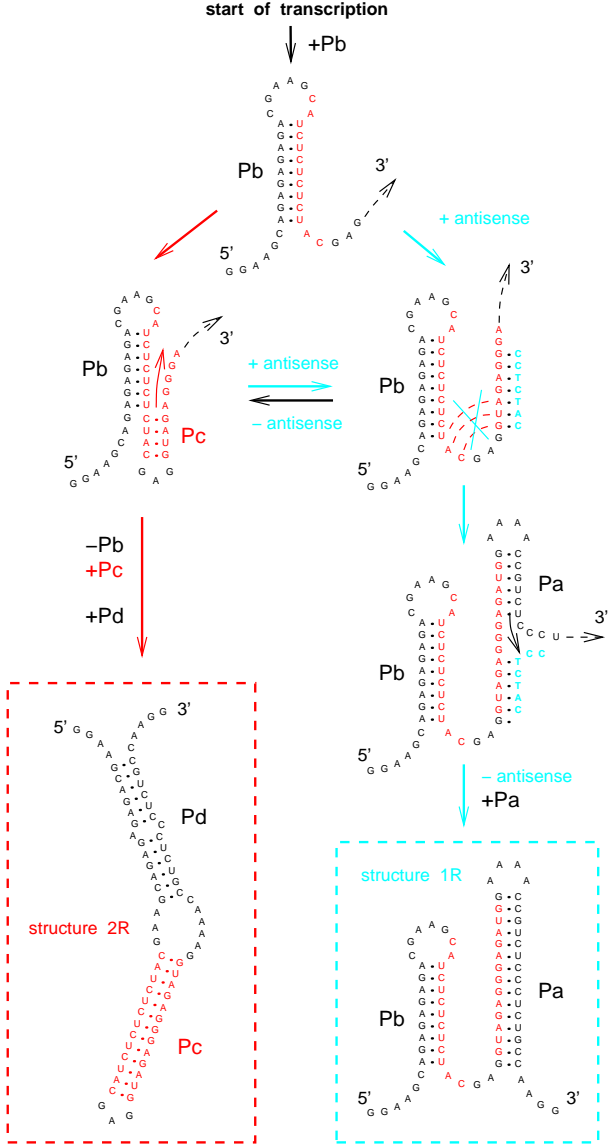}
\caption{\label{fig:wt} 
{\footnotesize {\bf Antisense regulation of co-transcriptional folding paths.} 
Interpretation of the encoded (left)
and redirected (right) co-transcriptional folding paths of the reverse switch
(Fig.~4). 
\textcolor [rgb]{0,0,0}{This is based on simulations performed using the
kinefold server\protect\cite{kinefold} 
({\tt  \footnotesize http://kinefold.curie.fr}); To simulate
the effect of antisense interaction, the 7mer and RNA switch sequences are
actually attached together via an inert linker (made of 'X' bases that do not
pair). }
}}
\vspace{-.5cm}
\end{figure}

\vspace{0.3cm}
\noindent
{\bf \em Bounding coding requirement through sequence symmetries}

\textcolor [rgb]{0,0,0}{In this subsection, we discuss how sequence symmetries
  can actually be used to estimate {\em necessary} base pairing conditions
  to encode efficient co-transcriptional
  folding paths. This requires, however, to reformulate base pairing
  conditions from an 
  {\em information content} perspective, following the approach developped for
  biomolecular sequences in refs\cite{adami,szostak}.}

In the following, we first establish a simple {\em conservation law} for information content.
We then argue that upperbounds for the coding requirement of selective folding paths (or other molecular features)
 can be estimated by restricting the available coding space with strong
 sequence  symmetries. 
Ultimately, upperbounds on coding requirements are related to the likelyhood
that a particular feature might arise from natural or {\it in vitro}
selection.  

\textcolor [rgb]{0,0,0}{Let us first recall what the information content of
a biomolecule is, before showing how it can actually be estimated for designed
RNA switches using sequence symmetries.} 

The information content $I$ of a functional biomolecule corresponds to
 the number of sequence constraints that have to be conserved to maintain its
 function under random mutations\cite{adami,szostak}. Expressed 
 in nucleotide unit, the maximum information content 
 that can be encoded on an $N$-nucleotide-long RNA sequence is precisely $I_{\rm max} = N$~nucleotides, which define a {\it unique} sequence
 amongst all $D^N$ different RNA sequences with $N$ nucleotides, where $D$ is the size of the coding alphabet ($D$=4 for nucleic acids; $D$=20
 for proteins). The fact that neutral mutations can accumulate on an RNA sequence without altering its function implies $I < I_{\rm max}=N$
 and can be simply translated into a {\em conservation law}, $I+J=N$, where
 $J$, the sequence entropy, corresponds to the number of unconstrained 
 nucleotides which generate $\Omega=D^J$ RNA sequences with the same function. Hence $J =\log_D(\Omega)$ and $I=N-J=\log_D(D^N) -\log_D(\Omega)$
 \cite{adami,szostak}. While $J$ and $I$ can be inferred by sampling sequence space as demonstrated in \cite{szostak}, their contributions to $N$
 do {\em not} usually correspond to a simple partition between $J$ ``meaningless'' and $I$ ``meaningful'' bases since many sequence constraints
 actually arise from non-local base-base correlations, as shown, in particular, with base pair covariations between homologous RNA sequences.
For this reason, 
  it is usually instructive to quantify information content separately within
 paired and unpaired regions,  when considering RNA structures.
  Since about 70\% of all bases are usually paired in low energy RNA structures, these base pairs typically contribute the most to total information
  content and thereby to the minimum coding requirement for a given RNA
 function. Hence, in the following, 
we will focus, for simplicity, on short RNA sequences ({\it e.g.} $N<$100 nucleotides) and
consider, {\it at first}, only base paired regions, ignoring both wobble base
pairs and unpaired regions. 

With these crude initial assumptions,
the information content of a short RNA sequence adopting a {\it unique} stable secondary structure can be estimated as $I \simeq J \simeq N/2$, since
the first base of each Watson-Crick base pair can be chosen arbitrarily. 
Hence, overall,
short RNA sequences adopting a {\it unique} stable secondary structure present a large sequence entropy $J=J_u \simeq N/2$ that can, in principle,
 be used to encode additional features such as alternative, low energy structures\cite{reidys,flamm} or possibly other molecular properties like co-transcriptional folding pathways, as shown in the previous subsection.
For example, encoding the simple bistable RNA of Fig.~1A but with exactly
overlapping helices
($\vert$Pa$\vert=\vert$Pb$\vert=\vert$Pc$\vert=\vert$Pd$\vert$) requires 
\textcolor [rgb]{0,0,0}{that $I\simeq 3N/4$ nucleotides be fixed} once the
initial $J=J_b\simeq N/4$ bases are 
chosen arbitrarily ({{\em e.g.}}~in the first pairing region). 
Including also
as stable structure the pseudoknot constructed around the same four
complementary regions (so as to obtain a tri-stable RNA molecule) then implies
that each pairing region is self-complementary and that only $J\simeq N/8$
bases can be chosen arbitrarily ({{\em e.g.}}~in the first half of the first
pairing region). Similarly, designing the same bistable RNA of Fig.~1A but with symmetrical helices (conserved under 5'-3' sequence inversion) implies that only around $J=J_{bs}\simeq N/8$ bases can be chosen arbitrarily ({{\em e.g.}}~in the first half of the first pairing region).

\textcolor [rgb]{0,0,0}{Similar  estimates can be made including 
wobble base pairs (GU and UG) in addition to 
Watson-Crick base pairs (GC, CG, AU and UA). In that
case, the available sequence entropy becomes}
\textcolor [rgb]{0,0,0}{ $J_u \simeq \log_4(6)\cdot N/2
\simeq 0.65N$  for a molecule with a {\em unique}
structure  ({\em i.e.} with 6 possible base pairs),  while  we get for the
previous  {\em bistable} RNA, $J_b \simeq \log_4(14) \cdot N/4 \simeq 0.48N$ 
({\em i.e.} with 14 possible quadruplet ``circuits'' including circular
permutations: $2\times ^{GC}_{CG}$, $2\times ^{AU}_{UA}$, $2\times 
^{GU}_{UG}$, $4\times ^{GU}_{CG}$ and $4\times ^{AU}_{UG}$) or $J_{bs} 
\simeq \log_4(14) \cdot N/8 \simeq 0.24N$ 
with additional {\em symmetric} helix restriction, as above.
Hence, the possibility of wobble pairs 
tends to increase sequence entropy $J$, and to concomitantly decrease 
the number of sequence constraints $I$. On the other hand, 
including a significant fraction of wobble pairs ({\em e.g.} a
third) in  designed structures tends also to facilitate
the formation of unwanted, alternative low 
energy structures ({\em e.g.} with fewer wobble pairs). 
Thus, including wobble and WC pairs  on an equal footing effectively 
{\em underestimates} coding requirements, since preventing the formation of
unwanted alternative structures then requires
additional information constraints, 
especially for longer sequences ({\it e.g.} $>$150 
nucleotides). 
In practice, limiting the fraction of wobble pairs in
designed structures can efficiently prevent the formation of alternative
structures with limited additional sequence constraint. 
({\em e.g.}~we have used 3 wobble pairs out of a total of 50 base pairs
in structures 1 and 2). 
This also justifies {\em a posteriori} the initial crude estimate we have made 
based on WC pairs only. Ignoring {\em opposite} effects of wobble base pairs 
and unwanted alternative structures is a reasonable first approximation
of {\em global} information constraint requirement. More precised estimates 
are difficult to obtain, in general, and sequence
candidates should always be tested for possible alternative structures
with an appropriate RNA folding algorithm including base pair stacking free
energies (we have used kinefold\cite{kinefold} which also includes pseudoknots
and knots in RNA structures). Alternatively,  it is usually
possible, owing to the available sequence entropy, to implement
highly constraining heuristics that prevent the
formation of alternative structures. Typical heuristics are based on
the limitation of short complementary substring occurrences in the
sequence\cite{seeman0,pierce,nanoletter}.}

\textcolor [rgb]{0,0,0}{The previous coding requirement estimates demonstrate
  that} structural information $I$ and symmetry constraints $I_s$ encoded on an RNA sequence can equivalently limit its available entropy $J$. In particular, combinations of structure and symmetry constraints can provide tighter upperbounds to possible coding increments $I'$ of any new feature encoded on the sequence, via the conservation law $I+I_s+J=N$, {\it i.e.,} $I'<J=N-I-I_s$.

This can be applied to estimate the minimum 
information that might be required to obtain two efficient opposite folding
paths from the generic bistable RNA switch sequence of Fig.~1A. 
From the previous estimate, we conclude that the {\it two} efficient
folding paths for the direct and reverse sequences do not
require to constrain more than about 1/8th of 
the 42 bases that are paired in {\it both} low energy structures ({\it i.e.}
`overlapping base pairs'). It corresponds to assigning, for each
folding path, a maximum of 2 or 3  {\it
  overlapping} base pairs not already constrained by the combination of 
branched and rod-like low energy structures. 

This limited coding requirement concerning overlapping base pairs reinforces,
{\it a  posteriori}, our intuitive design principles focussing, instead, on a
few bases paired in only one of the two low energy structures ({\it i.e.}
`non-overlapping base pairs'), Figs.~1\&2. Thus, efficient co-transcriptional
folding of the direct switch (Fig.~2) primarily relies on the {\it sole}
  terminal GC base pair at the base of Pa to prevent the nucleation of Pc,
  while the exchange between Pb and Pc for the reverse
  switch hinges on the AC terminal mismatch at the base of Pb to facilitate
  the nucleation of Pc. 

Hence, if all non-functional sequence symmetries
  are lifted, we expect that selective folding paths can
  indeed be readily achieved for a {\it wide class} of RNA sequences, as they
  require little encoded information beyond {\it small} asymmetries between
  alternative helices to guide or prevent their successive  exchanges during
  transcription. 
Interestingly, this pivotal role of a few unpaired or transiently paired bases for efficient folding paths is also observed for other encoded molecular functions of RNAs. For instance, a few unpaired conserved bases usually prove essential for ribozyme functions or {\it in vitro} selected aptamers showing remarkable binding efficiency to specific target molecules\cite{szostak}.

\begin{figure}
\includegraphics{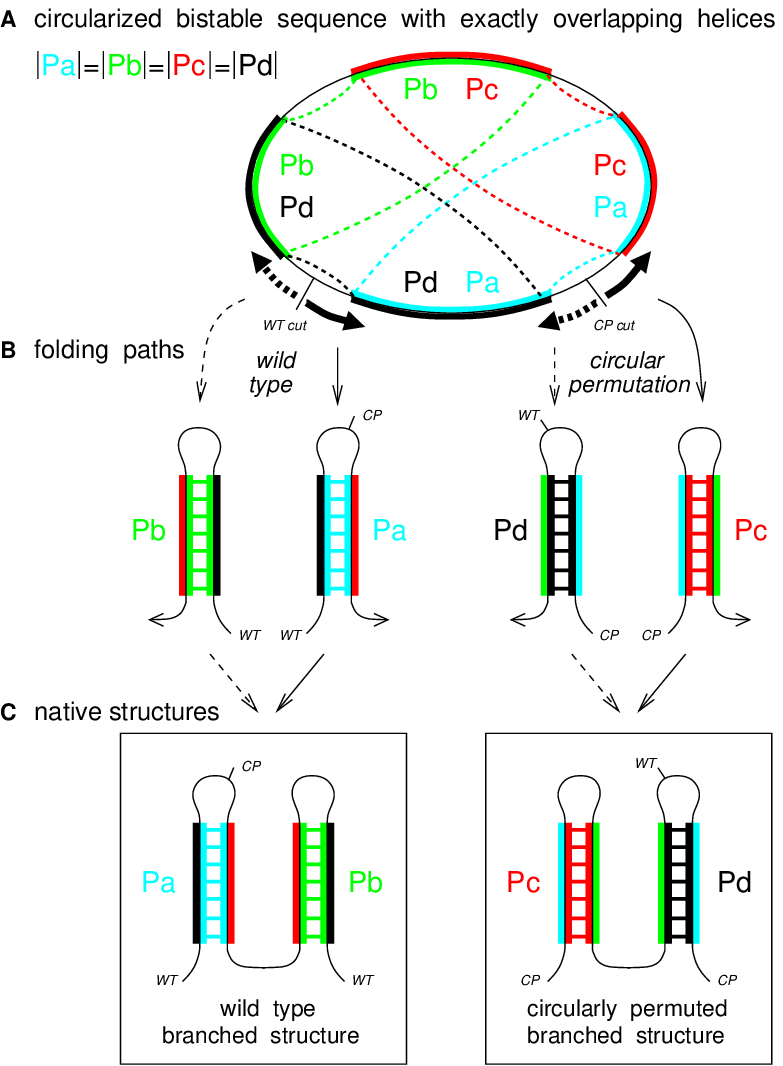}
\caption{\label{fig:wt} 
{\footnotesize
{\bf Simple sequential folding of a bistable RNA switch 
under sequence reversal and circular permutation.}\\ 
{\bf A.} Bistable generic sequence with exactly overlapping helices Pa, Pb, Pc
and Pd. A permutation between the starting and ending regions of the wild
type sequence can be obtained by genetically connecting its 5' and 3' ends and
engineering two new ends from an alternative break point in the 
circularized sequence. 
{\bf B.} Sequential folding path of the wild type vs circularly permuted
sequences. 
{\bf C.} Different branched structures obtained for wild type {\it and}
circularly permuted sequences independently from the direction of
transcription or sequence reversal (solid vs dashed arrows). 
The alternative rodlike structures (wild type Pd-Pc and 
circularly permuted Pa-Pb) are {\it not} formed through sequential folding,
although they are expected to coexist with branched structures at
equilibrium (not drawn).
Figs.~1\&2 
show, however, that small asymmetries (2-3 bases) between 
overlapping helices are sufficient to efficiently guide RNA switches into
either branched or  rod-like structures (see text).
}}
\end{figure}


\vspace{0.5cm}
\noindent
{\large \bf Discussion}

\vspace{0.2cm}
\noindent
{\bf \em Sequential vs encoded co-transcriptional folding}

Although many convincing reports have shown 
the importance of co-transcriptional folding\cite{repsilber,groeneveld,gultyaev1995,poot,gerdes1997,gultyaev1997,franch1997,nagel1999},
the underlying {\it mechanisms} behind
 {efficient} folding paths 
has remained elusive. As recalled in the introduction, this is mainly 
due to intrinsic difficulties to probe natural RNA
folding paths independently from their equilibrium structures and possibly
multiple functions.

In particular, studying the equilibrium folds of increasingly longer
3'-truncated transcripts has been argued to miss important out-of-equilibrium
intermediates on the folding path of full length molecules\cite{hdv}. 
This problem was, however, circumvented by  using circularly permutated
variants of wild type sequences to study the folding pathways of {\em E. coli}
RNaseP RNA\cite{pan1999,pan2} and {\em Tetrahymena} group I
intron\cite{woodson,heilman1}.
In this approach depicted with a simple
RNA switch on Fig.~6, the natural 5' and 3' ends of the 
molecule are genetically connected, while two new ends are engineered from an alternative break point in the
circularized sequence. This results in a circular permutation of the various domains on the linearized RNA. The
resulting permutation in their transcription order indirectly probes
folding paths of the full length wild type sequence. 
In particular, altering the connectivity of
 the primary sequence may lead to alternative co-transcriptional
 folds  as illustrated on Fig.~6. 
The rationale underlying such circular permutation scenario 
assumes that co-transcriptional folding primarily relies on 
sequential folding of the nascent chain into {\em independent} native domains
with no specific coding for transient base pair interactions during
transcription. 
This suggests, in particular, that co-transcriptional folding favors
branched secondary structures\cite{woodson}(as illustrated in Fig.~6) and
that the 5' to 3' 
directionality of RNA transcription plays little role as 
long as the different native domains can all successively fold during
transcription. 

By contrast, our results demonstrate that folding paths can 
efficiently  {\em guide} RNA transcripts into distinct alternative 
structures even when competing branched-like conformations exist and could, in
principle, form during transcription.
This competition between local overlapping helices and even global  
alternative structures is, in fact, ubiquitous to the folding dynamics and
thermodynamics of RNA molecules. 
For instance, co-transcriptional folding has long been known to induce
structural rearrangements as the nascent RNA 
chain is being transcribed\cite{kramer,nussinov}, demonstrating that transient 
helices almost inevitably
participate in co-transcriptional folding paths. 
More recently, transient helices were also shown to affect force-induced
unfolding 
paths of single RNA molecules in micromechanical experiments\cite{pullingepje}
({\em e.g.} for {\em E. coli}, 1540-nt-long 16S rRNA).

The present study, following earlier stochastic folding simulations
reported in refs.\cite{hdv,pnas2003,kinefold,pullingepje}, demonstrates 
that this unavoidable competition between alternative base pairs can 
be exploited to  precisely
encoded co-transcriptional folding paths of RNA sequences.
It is primarily achieved 
through branch migration exchanges between transient and native helices
forming successively as the transcription proceeds from 5' to 3' ends of the
sequence. 
This experimental finding is, in fact, corroborated by a recent 
statistical analysis of 
non-coding RNA sequences by Meyer \& Mikl\'os\cite{meyer} who demonstrated 
the existence of a
5'-3' vs 3'-5' asymmetry in the relative positional correlation between native
and non-native helices along primary sequences.

\vspace{0.3cm}
\noindent
{\bf \em Information content and RNA evolution}

Non-coding RNAs typically tolerate a significant number of neutral mutations and covariations in their sequence,
which presumably facilitates their continuous adaptation to environmental changes. From an information content perspective,
this tolerance to (concerted) mutations also suggests that (partly) unconstrainted nucleodites may be used to encode other
alternative structures and functions on the same RNA sequence, a feature which might have favored the emergence of new functional
 RNAs and RNA switches in the course of evolution\cite{schultes,schuster}.
In fact, {\it much more} functional, as well as non-functional information can be encoded on an RNA sequence. For instance, the pair
of RNA switches we have designed (Fig.~2) demonstrates that not only alternative structures but also selective folding
pathways {\it and} strong sequence symmetries can all be encoded {\it simultaneously} on the same RNA sequence. While such strong sequence
symmetries are both non-functional and probably too stringent constraints to possibly emerge and adapt through natural or {\it in vitro}
selection, they can be used to provide quantitative information on other
encoded features of interest.
For instance, equalling stacking contributions between alternative folds using
helix symmetries also provides a powerful {\it differential} approach to
uncover other free energy contributions 
from non-canonical tertiary structure motifs. 
In the present study, small but
reproducible differences in band separation between direct and reverse
switches at equilibrium (Fig.~2) may possibly reflect a difference in tilt
angle between helices Pc and Pd, due to differently structured interior loops 
in their respective rod-like structures. 

In this study, we showed that co-transcriptional folding can efficiently
guide RNA  
folding either towards branched structures (as for the `direct' switch, Fig.~2)
or towards elongated rod-like structures (as for the `reverse' switch, Fig.~2)
even though their helices are mutually identical. We also argued that only
limited information is necessary to encode such selective
folding paths for generic RNA switches: it essentially amounts to encoding the
relative lengths of helices forming successively during transcription. 
Moreover, this strict hierarchy between successively exchanging helices can
be somewhat alleviated by resorting to  topological barriers based on
`entangled' helices ({\em i.e.} simple co-transcriptional
knots)\cite{kinefold}. Hence,  we expect that the present findings concerning
short RNA sequences ({\em i.e.} $<$100~bases) may be applied to design
efficient folding paths for a wide class of 
larger RNA target structures.
This could be achieved by encoding different series of local
folding events leading to a succession of either rod-like or branched motifs
at the 3' end of the nascent RNA molecule during transcription. 
Such folding scheme also provides a theoretical frame to analyze selective
folding paths of natural non-coding RNAs\cite{hdv}. 

Finally, these results suggest that efficient folding pathways might 
have easily emerged and continuously adapted in the course of evolution the
same way functional native 
structures have done so through mutation drift in sequence space; 
non-deleterious mutations are mostly neutral and conserve sequence folds and
activity, while new functions may occasionally arise by rare hopping  
between intersecting networks of neutral mutations (``neutral
networks'')\cite{reidys,schultes,schuster}. 
Furthermore, the fact that encoded folding paths may be redirected through
transient antisense interactions 
(Figs.~4,5) provides simple `all-RNA' mechanisms to regulate the functional
folding of RNAs in the absence of any elaborate control at the level of
transcription initiation. 
From an ancestral ``RNA World'' perspective, this {\em constitutive} coupling
between RNA synthesis and RNA folding regulation may have also enabled the
early emergence of autonomous RNA-based networks relying solely on intra- and
intermolecular base pair interactions. Indeed,  RNA molecules
cross-regulating their respective encoded folding paths 
could, in principle, be combined to perform
essential regulation tasks, characteristic to all natural and engineered
control networks ({\em e.g.} negative and positive feedback loops, feedforward
loops, toggle switches, oscillators, etc).

\vskip 0.5cm

\noindent
{\bf Acknowledgements} \\
We thank H. Putzer and L. Hirschbein for critical reading
of the manuscript and R. Breaker, D. Chatenay, B. Masquida, K. Pleij,
P. Schuster, J. Robert and S. Woodson for discussions. We 
acknowledge support from CNRS, Institut Curie, Minist\`ere de la Recherche
(ACI grant \#DRAB04/117) and HFSP grant (\# RGP36/2005).

\vfill
\eject


\begin{thebibliography}{99}

\footnotesize 


\bibitem{dahlberg} Dahlberg, A.E. (2001) 
The Ribosome in Action.
{\em Science}, {\bf 292}, 868-869.

\bibitem{cech} Kruger, K., Grabowski, P., Zaug, A.J., Sands, J., Gottschling,
 D.E. \& Cech, T.R. (1982) 
Self-splicing RNA: Autoexcision and autocyclization of the ribosomal RNA intervening sequence of Tetrahymena. 
 {\em Cell} {\bf 31}, 147-157.

\bibitem{bartelszostak} Bartel, D.P. \& Szostak, J.W. (1993) 
Isolation of new ribozymes from a large pool of random sequences. 
{\em Science} {\bf 261}, 1411-1418.

\bibitem{joyce} Joyce, G.F. (1989) 
Amplification, mutation and selection of catalytic RNA. 
{\em Gene} {\bf 82}, 83-87. 

\bibitem{ellington} Ellington, A.E. \& Szostak, J.W. (1990) 
In vitro Selection of RNA Molecules that Bind Specific Ligands. 
{\em Nature} {\bf 346}, 818-822.

\bibitem{tuerk} Tuerk, C. \& Gold, L. (1990) 
Systematic evolution of ligands by exponential enrichment: RNA ligands to bacteriophage T4 DNA polymerase. 
{\em Science} {\bf 249}, 505-510.

\bibitem{mironov} Mironov, A.S., Gusarov, I., Rafikov, R., Lopez, L.E.,
 Shatalin, K., Kreneva, R.A., Perumov, D.A. \& Nudler E.  (2002)
Sensing small molecules by nascent RNA: a mechanism to control transcription in bacteria.
{\em Cell} {\bf 111}, 747-756.

\bibitem{nahvi} Nahvi, A., Sudarsan, N., Ebert, M.S., Zou, X., Brown, K.L. \&
 Breaker, R.R. (2002)
Genetic control by metabolite binding mRNA.
{\em Chem. Biol.} {\bf 9}, 1043-1049.

\bibitem{winkler} Winkler, W.C. \& Breaker, R.R. (2003)
Genetic control by metabolite-binding riboswitches.
{\em ChemBioChem} {\bf 4}, 1024-1032.

\bibitem{bartel} Bartel, D.P. (2004)
MicroRNAs: genomics, biogenesis, mechanism and function.
{\em Cell} {\bf 116}, 281-297.

\bibitem{nudler} Nudler, E. \& Mironov, A.S. (2004)
The riboswitch control of bacterial metabolism.
{\em Trends Biochem. Sci.} {\bf 29}, 11-17.

\bibitem{hannon} He, L. \& Hannon, G.J. (2004)
MicroRNAs: small RNAs with a big role in gene regulation.
{\em Nature Reviews Genetics} {\bf 5}, 522-531.

\bibitem{breaker} Winkler, W. C., Nahvi, A., Roth, A., Collins, J. A. \&
 Breaker, R. R. (2004)
{\em Nature} {\bf 428}, 281-286.

\bibitem{uhlenbeck} Uhlenbeck O. C. (1995)  
Keeping RNA happy.
{\em RNA} {\bf 1}, 4-6.

\bibitem{boyle} Boyle, J., Robillard, G. \& Kim, S. (1980)
Sequential folding of transfer RNA. A nuclear magnetic resonance study of successively longer tRNA fragments with a common 5'end.
{\em J. Mol. Biol.} {\bf 139}, 601-625.

\bibitem{kramer} Kramer, F. \& Mills, D. (1981)
Secondary structure formation during RNA synthesis.
{\em Nucleic Acids Res.} {\bf 9}(19), 5109-5124.

\bibitem{nussinov} Nussinov, R. \& Tinoco, I. Jr. (1981)
Sequential folding of a messenger RNA molecule.
{\em J. Mol. Biol.} {\bf 151}(3), 519-533.


\bibitem{biebricher} Biebricher, C.K. \& Luce, R. (1992)
In vitro recombination and terminal elongation of RNA by Q$\beta$ replicase.
{\em EMBO J.} {\bf 11}(13), 5129-5135.

\bibitem{crothers} LeCuyer, K. A. \& Crothers, D. M.
Kinetics of an RNA conformational switch 
{\em Proc. Natl. Acad. Sci. USA}, {\bf 91}, 333-3377 (1994).

\bibitem{gultyaev1998} Gultyaev, A.P., van Batenburg, F.H. \& Pleij,
 C.W. (1998)
Dynamic competition between alternative structures in viroid RNAs simulated by an RNA folding algorithm.
{\em J. Mol. Biol.} {\bf 276}(1), 43-55.

\bibitem{repsilber} Repsilber, D., Wiese, S., Rachen, M., Schroder, A.,
 Riesner, D. \& Steger, G. (1999)
Formation of metastable RNA structures by sequential folding during transcription: Time-resolved structural analysis of potato spindle tuber viroid (-)-stranded RNA by temperature-gradient gel electrophoresis.
 {\em RNA} {\bf 5}, 574-584.

\bibitem{olsthoorn} Olsthoorn, R.C., Mertens, S., Brederode, F.T. \& Bol,
 J.F. (1999)
A conformational switch at the 3' end of a plant virus RNA regulates viral replication.
{\em EMBO J.} {\bf 18}(17), 4856-4864.

\bibitem{putzer1} Putzer, H., Gendron, N. \& Grunberg-Manago, M. (1992)
Coordinate ex-pression of the two threonyl-tRNA synthetase genes in Bacillus subtilis: Control by transcriptional antitermination involving a con-served regulatory sequence
{\em EMBO J.} {\bf 11}, 3117-3127.

\bibitem{putzer} Putzer, H., Condon, C., Brechemier-Baey, D., Brito, R. \&
 Grunberg-Manago, M. (2002)
Transfer RNA-mediated antitermination in vitro.
{\em Nucleic Acids Res.} {\bf 30}(14), 3026-3033.

\bibitem{wickiser} Wickiser, J.K., Winkler, W.C., Breaker, R.R. \& Crothers,
  D.M. (2005)
The Speed of RNA Transcription and Metabolite Binding Kinetics Operate an FMN Riboswitch.
{\em Mol. Cell.} {\bf 18}, 49-60. 

\bibitem{romby} Romby, P. \& Springer, M. (2003)
Bacterial translational control at atomic resolution.
{\em Trends Genet.} {\bf 19}(3), 155-161. Review.

\bibitem{caillet} Caillet, J., Nogueira, T., Masquida, B., Winter, F., Graffe,
 M., Dock-Bregeon, A.C., Torres-Larios, A., Sankaranarayanan, R., Westhof,
 E., Ehresmann, B., Ehresmann, C., Romby, P. \& Springer, M. (2003)
The modular structure of Escherichia coli threonyl-tRNA synthetase as both an enzyme and a regulator of gene expression.
{\em Mol. Microbiol.} {\bf 47}(4), 961-974.

\bibitem{brunel2} Brunel, C., Romby, P., Sacerdot, C., de Smit, M., Graffe,
 M., Dondon, J., van Duin, J., Ehresmann, B., Ehresmann, C. \& Springer, M.
 (1995)
Stabilised secondary structure at a ribosomal binding site enhances translational repression in E. coli.
{\em J. Mol. Biol.} {\bf 253}(2), 277-290.

\bibitem{moller-jensen} Moller-Jensen, J., Franch, T. \& Gerdes, K. (2001)
Temporal translational control by a metastable RNA structure.
{\em J. Biol. Chem.} {\bf 276}(38), 35707-35713.

\bibitem{vanmeerten} van Meerten, D., Girard, G. \& van Duin, J. (2001)
Translational control by delayed RNAfolding: identification of the kinetic trap.
{\em RNA}. {\bf 7}, 483-494.

\bibitem{smit} de Smit, M.H. \& van Duin, J. (2003)
Translational standby sites: how ribosomes may deal with the rapid folding kinetics of mRNA.
{\em J. Mol. Biol.} {\bf 331}(4), 737-743.

\bibitem{soukup} Soukup, G.A. \& Breaker, R.R. (1999)
Engineering precision RNA molecular switches
{\em Proc. Natl. Acad. Sci. USA} {\bf 96}, 3584-3589.

\bibitem{nagel2002} Nagel, J.H. \& Pleij, C.W. (2002)
Self-induced structural switches in RNA.
{\em Biochimie.} {\bf 84}(9), 913-923. Review.



\bibitem{groeneveld} Groeneveld, H., Thimon, K. \& van Duin, J. (1995)
Translational control of maturation-protein synthesis in phage MS2: a role for the kinetics of RNA folding?
{\em RNA} {\bf 1}(1), 79-88.

\bibitem{gultyaev1995} Gultyaev, A.P., van Batenburg, F.H. \& Pleij, C.W.
 (1995)
The influence of a metastable structure in plasmid primer RNA on antisense RNA binding kinetics.
{\em Nucleic Acids Res.} {\bf 23}(18), 3718-3725.

\bibitem{poot} Poot, R.A., Tsareva, N.V., Boni, I.V. \& van Duin, J. (1997)
RNA folding kinetics regulates translation of phage MS2 maturation gene.
{\em Proc. Natl. Acad. Sci. USA} {\bf 94}, 10110-10115.

\bibitem{gerdes1997} Gerdes, K., Gultyaev, A.P., Franch, T., Pedersen, K. \&
 Mikkelsen, N.D. (1997)
Antisense RNA-regulated programmed cell death.
{\em Annu. Rev. Genet.} {\bf 31}, 1-31. Review.

\bibitem{gultyaev1997} Gultyaev, A.P., Franch, T. \& Gerdes, K. (1997)
Programmed cell death by hok/sok of plasmid R1: coupled nucleotide covariations reveal a phylogenetically conserved folding pathway in the hok family of mRNAs.
{\em J. Mol. Biol.} {\bf 273}(1), 26-37.

\bibitem{franch1997} Franch, T., Gultyaev, A.P. \& Gerdes, K. (1997)
Programmed cell death by hok/sok of plasmid R1: processing at the hok mRNA 3'-end triggers structural rearrangements that allow translation and antisense RNA binding.
{\em J. Mol. Biol.} {\bf 273}(1), 38-51.

\bibitem{nagel1999} Nagel, J.H., Gultyaev, A.P., Gerdes, K. \& Pleij,
 C.W. (1999)
Metastable structures and refolding kinetics in hok mRNA of plasmid R1.
{\em RNA} {\bf 5}(11), 1408-1418.


\bibitem{pan1999} Pan, T., Fang, X. \& Sosnick, T. (1999)
Pathway modulation, circular permutation and rapid RNA folding under kinetic control.
{\em J. Mol. Biol.} {\bf 286}, 721-731.

\bibitem{pan2} Pan, T., Artsimovitch, I., Fang, X., Landick, R. \& Sosnick,
T.R. (1999)
Biochemistry Folding of a large ribozyme during transcription and the effect of the elongation factor NusA.
{\em Proc. Natl. Acad. Sci. USA} {\bf 96}, 9545-9550.

\bibitem{woodson} Heilman-Miller, S.L. \& Woodson, S.A. (2003)
Perturbed folding kinetics of circularly permuted RNAs with altered topology.
{\em J. Mol. Biol.} {\bf 328}(2), 385-394.

\bibitem{heilman1} Heilman-Miller, S.L. \& Woodson, S.A. (2003)
 Effect of transcription on folding of the Tetrahymena ribozyme.
{\em RNA}. {\bf 9}(6), 722-733.

\bibitem{pnas2003} Xayaphoummine, A., Bucher, T., Thalmann, F. \& Isambert,
 H. (2003)
 Prediction and statistics of pseudoknots in RNA structures using exactly
 clustered stochastic simulations. 
 {\em Proc. Natl. Acad. Sci. USA}, {\bf 100}, 15310-15315.

\bibitem{kinefold} Xayaphoummine, A., Bucher, T. \& Isambert, H.  (2005)  
Kinefold web server for RNA/DNA folding path and structure prediction
   including pseudoknots and knots.  
{\em Nucleic Acids Res.} {\bf 33}, 605-610.

\bibitem{adami} Adami, C. \& Cerf, N.J. (2000)
Physical complexity of symbolic sequences.
{\em Physica D} {\bf 137}, 62-69.

\bibitem{szostak} Carothers, J.M., Oestreich, S.C., Davis, J.H. \& Szostak,
 J.W. (2004)
Informational complexity and functional activity of RNA structures.
{\em J. Am. Chem. Soc.} {\bf 126}, 5130-5137.

\bibitem{nanomemories} Isambert, H. 
Voltage addressable nanomemories in DNA?
{\em C. R. Physique}, {\bf 3}, 391-396 (2002).

\bibitem{chamberlin}
Chamberlin, M. \& Ring, J. (1973) 
Characterization of T7-specific ribonucleic acid polymerase.
{\em J. Biol. Chem.} {\bf 248}, 2235-2244; {\bf 248}, 2245-2250.

\bibitem{nanoletter}  
 Viasnoff, V. , Meller, A., \&  Isambert, H. (2006)
DNA nanomechanical switches under folding kinetics control,
{\em Nano Lett.} {\bf 6}, 101-104.

\bibitem{seeman0} Seeman, N.C.  (1982)
Nucleic acid junctions and lattices. 
{\em J. Theor. Biol.} {\bf 99}, 237-247.

\bibitem{pierce}
Dirks, R.M., Lin, M., Winfree, E. \& Pierce, N.A. (2004)
Paradigms for computational nucleic acid design.
{\em Nucleic Acids Res.} {\bf 32}, 1392-1403.

\bibitem{reidys} Reidys, C.M., Stadler, P.F. \& Schuster, P. (1997)
Generic properties of combinatory maps: Neural networks of RNA secondary
structures
{\em Bull. Math. Biol.}, {\bf 59}, 339-397.

\bibitem{flamm} Flamm, C., Hofacker, I. L., Maurer-Stroh, S., Stadler,
 P. F. \& Zehl, M. (2001)
Design of multistable RNA molecules.
{\em RNA}, {\bf 7}, 254-265.


\bibitem{hdv} Isambert, H. \& Siggia, E.D. (2000)
Modeling RNA folding paths with pseudoknots: application to hepatitis delta virus ribozyme.
{\em Proc. Natl. Acad. Sci. USA}, {\bf 97}, 6515.

\bibitem{pullingepje} Harlepp, S., Marchal, T., Robert, J., L\'eger, J.-F.,
Xayaphoummine, A., Isambert, H. \& Chatenay, D. (2003)
Probing complex RNA structures by mechanical force.
{\em Eur. Phys. J. E}, {\bf 12}, 605-615.

\bibitem{meyer} Meyer, I.M. \& Mikl\'os, I. (2004)
Co-transcriptional folding is encoded within RNA genes.
{\em BMC Molecular Biology} {\bf 5}, 10.

\bibitem{schultes} Schultes, E.A. \& Bartel, D.P. (2000)
One sequence, two ribozymes: implications for the emergence of new ribozyme
folds. 
{\em Science} {\bf 289}, 448-452.

\bibitem{schuster}
Schuster, P., Fontana,  W., Stadler,  P. F. \& Hofacker, I. L.  (1994)
{\em Proc. R. Soc. London B} {\bf 255}, 279. 

\end{thebibliography}
\end{document}